# Substantial enhancement in thermoelectric figure-of-merit of half Heusler ZrNiPb alloys

Amardeep Sagar[#1], Aman Bhardwaj[#1], Andrei Novitskii[2], Vladimir Khovaylo[2,3] and Satyabrata Patnaik[*1]

[1]School of Physical Sciences, Jawaharlal Nehru University, New Delhi-110067, India
[2]National University of Science and Technology "MISiS" Moscow 119049, Russia
[3]Belgorod State University, Belgorod 308015, Russia
[*]Corresponding author: spatnaik@mail.jnu.ac.in
[#]Both the authors contributed equally

**Abstract:**

Ternary half Heusler alloys are under intense investigations recently towards achieving high thermoelectric figure-of-merit (ZT). Of particular interest is the ZrNiPb based half Heusler (HH) alloy where an optimal value of ZT ~ 0.7 at 773 K has been achieved by co-doping Sn and Bi at Pb site. In this work, we identify an excellent ZT of 1.3 in $ZrNi_{1+x}Pb_{0.38}Sn_{0.6}Bi_{0.02}$ (x= 0.03, at 773 K) composite alloy. This is achieved by synergistic modulation of electronic as well as thermal properties via introduction of minor phase of full Heusler (FH) in the HH matrix through compositional tuning approach. These Ni-rich $ZrNi_{1+x}Pb_{0.38}Sn_{0.6}Bi_{0.02}$ ($0 \leq x \leq 0.07$) alloys were synthesized via Arc melting followed by consolidation via Spark Plasma Sintering (SPS). These alloys were characterized by XRD and SEM that shows formation of nanocomposites comprising of HH matrix phase and FH secondary minor phases. Enhancement in ZT is mainly attributed to a synchronized increase in power factor (~42%) and about ~ 25% decrease in its thermal conductivity. The thermoelectric compatibility factor (S) is also calculated for all samples. The value of $|S| \sim 2.7$ $V^{-1}$ (at 773K) is observed for x=0.03 which is ~17% higher than bare HH composition (x=0.0). The theoretically calculated thermoelectric device efficiency of best performing sample $ZrNi_{1.03}Pb_{0.38}Sn_{0.6}Bi_{0.02}$ is estimated to be $\eta \sim 13.6\%$. Our results imply that controlled fine tuning in HH compounds through compositional tuning approach would lead to novel off-stoichiometric HH phases with enhanced ZT value for efficient thermoelectric device fabrication.

1. Introduction

Thermoelectric (TE) materials and devices hold great promise as alternative sources of energy for garnering electricity from waste heat.[1] However, large scale commercialization of TE materials has remained limited due to poor conversion efficiency and high cost. The conversion efficiency is assessed from the thermoelectric figure-of-merit $ZT = \alpha^2\sigma T/\kappa$, where $\kappa$ is total thermal conductivity, $\sigma$ is electrical conductivity; T is the temperature (in Kelvin) and $\alpha$ is Seebeck coefficient. Due to interdependence of all the three parameters, enhancement in ZT has been a challenging task.[2] Along with the ZT optimization, estimation of thermoelectric compatibility factor (S) is an important parameter which is useful in finding compatible counterparts to attain maximum thermoelectric device conversion efficiency (η). The conversion efficiency of TE devices depends on the ZT value at operating temperature and the theoretical conversion efficiency of TE device (η) which is related to hot temperature ($T_h$), cold temperature ($T_c$) and the average value of ZT ($ZT_{avg}$).[3,4] With the advent of new theoretical paradigm and significant advancement in processing technologies, several new pathways have been attained to decouple the thermoelectric parameters and thereby attain high figure of merit and compatibility factor.[3,4]

Half-Heusler alloys (HH) are among the emerging TE materials for varied applications, particularly in the mid-temperature (400 °C to 800 °C) range. The main advantages of HH alloys include high mechanical strength, moderately good power factor (PF = $\alpha^2\sigma$) and high thermal stability for TE device fabrication. However, high thermal conductivity results in low ZT value in these materials.[5] In particular, n-type HHs such as ternary intermetallic (Ti/Zr/Hf)NiSn are considered potential candidates for waste heat recovery applications[6], but their reported ZT value is still low (~0.3 at 773 K). Besides, the use of costly Hafnium makes them expensive for industrial application.[7] Further possibility in enhancement of ZT for the (Ti/Zr/Hf)NiSn based materials seems daunting as several approaches have been tried yet their thermal conductivity continues to be high due to the effect of bipolar conduction at higher temperature.[8] Thus, the search for a new n-type HH material with high conversion efficiency has attracted considerable attention in the recent past.

In general terms, the thermal conductivity κ has two main components that are derived from lattice ($κ_L$) as well as electronic ($κ_e$) contributions. The approach related to nanostructuring/nanocomposite has recently been established to be of substantial success in improving ZT due to a large reduction in lattice $κ_L$. The microscopic origin of such phenomena involving phonon scattering are reasonably well understood.[9,10] However, the overall increase in ZT attained thus far remains impractical for technological applications. Clearly, there is a need to couple this approach with clever enhancement in α without altering σ. Hence a new strategy has been envisaged, in which nanocomposites derived by compositional tuning need to be optimized for a reasonable decrease in $κ_L$ along with simultaneous increase in Seebeck co-efficient.[11-15] Generally, the $κ_L$ value is dependent on the crystal lattice and microstructure. Therefore, a compositional tuning in HH materials can play a vital role in reducing the $κ_L$. Several recent papers have reported that formation of full Heusler ((Ti/Zr/Hf)Ni$_2$Sn) inclusion phase in half Huesler ((Ti/Zr/Hf)NiSn) matrix can lead to the desired result.[11-15] Both FH and HH have FCC structure, which helps to create *in-situ* formation of nanocomposites leading to lowering of $κ_L$. Furthermore, it is reported that the HH/FH band alignment leads to optimized carrier mobility for higher electrical conductivity. Most importantly such compositional tuning approaches have also been reported to increase Seebeck coefficient via intra-matrix electronic structure modifications. It is thus indicated that all three parameters of ZT dependence can be optimized simultaneously by following compositional tuning of HH alloys.[16,17]

Of great current interest is the HH alloys based on ZrNiPb system where an optimized high ZT ~ 0.75 (at 873 K) has been reported via co-doping of Sn and Bi.[18-20] This enhancement in ZT is primarily assigned to the alloying effect of Sn on Pb which results in a significant reduction in $κ_L$ while Bi doping tunes the carrier concentration for maintaining high power factor. The question is whether it can be further increased by following compositional variation techniques as implemented in MNi$_{1+x}$Sn (M=Ti/Zr/Hf) systems.[11-15] In such nanocomposite systems the inclusion of precipitated FH phase unto HH matrix is caused by phase segregation processes. In such nanocomposites reduction in $κ_L$ value can be attributed to various kind of scattering mechanism which includes grain boundary scattering, point scattering and alloy scattering.[11-15]Furthermore, the improved electronic properties observed in these nanocomposites due to compositional tuning leads to band-structure engineering which has been successfully demonstrated as an efficient approach for achieving high ZT of many TE materials.[21] With this evidence in mind, we intuitively understand that synthesis of an off-stoichiometric ZrNiPb based HH/FH nanocomposite material could possibly lead to higher ZT value.

Here, in this work, we carry out detailed synthesis and characterization of FH & HH nanocomposite to test the above hypothesis. We note that ZrNiPb and ZrNi$_2$Pb are the corresponding half Heusler and full Heusler components. Further, in order to obtain a nanocomposite of HH phase having FH inclusions with multiple length scales, Ni concentration was varied in ZrNi$_{1+x}$Pb$_{0.38}$Sn$_{0.6}$Bi$_{0.02}$ ($0 \leq x \leq 0.07$) compositions. We observe a large decrease in $\kappa_L$ (~25% for ZrNi$_{1.03}$Pb$_{0.38}$Sn$_{0.6}$Bi$_{0.02}$) which has been achieved due to tuning of various scattering mechanism caused by large difference in atomic mass, embedded FH inclusion and HH/FH interface. Also a substantially improved PF (~42% for composition; x=0.03) has been obtained. This is due to structural similarities between HH and FH which create HH/FH interface for optimizing the carrier mobility. An excellent value of ZT = 1.3 at 773 K has been attained for ZrNi$_{1.03}$Pb$_{0.38}$Sn$_{0.6}$Bi$_{0.02}$ which is highest ZT value in ZrNiPb based HH compositions.

## 2. Experiments

Towards synthesizing the materials, high purity lead (Pb; 99.99%, Alfa Aesar, granules), nickel (Ni; 99.8%, Alfa Aesar, powder), tin (Sn; 98.0%, CDH shots), zirconium, (Zr; 98.0%, CDH, powder), and bismuth (Bi; 99.99%, Alfa Aesar shots) were weighed in their respective stoichiometric ratio followed by arc melting under argon atmosphere. Here, 2 at. wt. % additional Pb was weighed in each composition because in arc melting process Pb gets partially evaporated. The ingots were melted three times followed by annealing in sealed quartz tube at 1073K for 7 days to obtain the phase stability. These annealed ingots were then subjected to mortal-pastle grinding followed by ball milling to get fine powder with reduced grain size. The powders were consolidated into highly dense pellets by using spark plasma sintering (SPS) technique at 800 °C and 50 MPa under high vacuum with a 10 minutes holding time in a graphite die of 12.7 mm diameter. A high vacuum is provided in SPS to achieve the clean samples by avoiding the adsorptive gases and impurities which thereby prevent the oxidation of the sample. The dense pellets obtained of samples by SPS were further annealed at 750 °C for 2 days. The densities of as-grown samples were observed to be 97% of the theoretical density. As discussed already, the excess Ni effectively leads to synthesis of nano-composites of FH and HH components. This can be represented by the following chemical equation:

$$Zr + (1+x)Ni + 0.60Sn + 0.38Pb + 0.02Bi \rightarrow (1-x)ZrNiSn_{0.6}Pb_{0.38}Bi_{0.02}(HH) + xZrNi_2Sn_{0.6}Pb_{0.38}Bi_{0.02}(FH) \qquad (1)$$

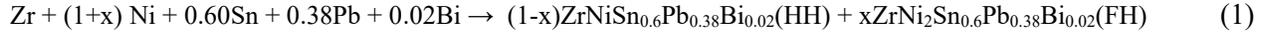

Structural characterization of the samples were done by X-ray Diffraction technique (Rigaku Mini flex II) for the phase identification of $ZrNi_{1+x}Pb_{0.38}Sn_{0.6}Bi_{0.02}$ ($0 \leq x \leq 0.07$) samples. The microstructural characterization on the $ZrNi_{1+x}Pb_{0.38}Sn_{0.6}Bi_{0.02}$ alloys was performed using Scanning Electron Microscopy (SEM). The actual atomic composition of alloys was estimated through Energy Dispersive Spectroscopy (EDAX). The thermal diffusivities ($D$) of all the synthesized samples were measured by using Laser flash apparatus (Lineseis; LFA 1000). Specific heat ($C_p$) along with density ($d$) was measured for all synthesized samples using differential scanning calorimetry (DSC) instrument and Archimedes principle, respectively. The thermal conductivity was calculated using the relation $\kappa = D \times C_p \times d$. The electronic transport properties were measured by the ULVAC ZEM-3 instrument. Room temperature carrier concentration ($n$) and mobility ($\mu$) were determined using the Hall Effect measurement system in conjunction with a superconducting magnet.

## 3. Results and Discussions

### 3.1. Structural Characterization

Figure 1 displays the XRD of $ZrNi_{1+x}Pb_{0.38}Sn_{0.6}Bi_{0.02}$ ($0 \leq x \leq 0.07$) composite samples. XRD of $ZrNiPb_{0.38}Sn_{0.6}Bi_{0.02}$ shows all the peaks of HH (space group F4-3m; no. 216; JCPDS card no. 00-023-1281) with a small trace of Pb as an impurity phase. The Pb impurities are expected as all the samples are vacuum sealed in quartz tube before annealing, and hence Pb becomes vapor during the annealing treatment and remains unreacted to small extent. Importantly, the XRD of $ZrNi_{1+x}Pb_{0.38}Sn_{0.6}Bi_{0.02}$ ($0 \leq x \leq 0.07$) composite samples clearly reveals the presence of FH (Fm3m; no. 225; JCPDS card no. 00-023-1282) phases along with HH and Pb impurity phase. Post arc melting, liquid melt of $ZrNi_{1+x}Pb_{0.38}Sn_{0.6}Bi_{0.02}$ when cooled down, crystallizes to form stable composite consisting of HH and FH components through phase separation.[22-24] Herein, the segregation of the FH and HH phases are produced due to excess concentration of Ni concentration.[22-24]

The scanning electron microscopy (SEM) was performed in several samples but in Fig. 2, we focus on ZrNi$_{1.03}$Pb$_{0.38}$Sn$_{0.6}$Bi$_{0.02}$ that was identified with best ZT for the entire series. The micrograph reveals a tightly packed microstructure of HH phase along with *in-situ* FH precipitates. The EDX of the FH (white dotted circle) and HH (solid white square) regions confirm their composition. These results provide direct evidence of the formation of FH precipitates as a secondary phase in the HH matrix.

**3.2 Electronic Transport Properties**

Figure 3(a) and 3 (b) plot Seebeck Coefficient and electrical conductivity as a function of temperature for all synthesized compositions. Here, all samples have negative value of α which imply that these samples are n-type semiconductors. The α value of ZrNi$_{1.03}$Pb$_{0.38}$Sn$_{0.6}$Bi$_{0.02}$ is ~ -124.5 µV/K at 323 K which is ~24% improved in contrast to ZrNiPb$_{0.38}$Sn$_{0.6}$Bi$_{0.02}$ HH (~ -100.7 µV/K at 323 K). More importantly, an increase of ~12% was achieved in α at higher temperature of 773 K. We further note that for composite with x= 0.07, the value of α was noted to be lower (i.e. α ~ -139.7 µV/K at 773K) as compared to the ZrNiPb$_{0.38}$Sn$_{0.6}$Bi$_{0.02}$ (HH; α ~ -161.9 µV/K at 773K) that clearly reveals that composite had crossed over to greater metallicity as will be evident in the following discussion.

An enhanced value of α for ZrNi$_{1.01}$Pb$_{0.38}$Sn$_{0.6}$Bi$_{0.02}$ and ZrNi$_{1.03}$Pb$_{0.38}$Sn$_{0.6}$Bi$_{0.02}$ as compared to the HH alloy was obtained due to reduced value of *n* at 323K (i.e. $\alpha = \frac{8\pi^2 K_B^2}{3eh^2} m_e^* T \left(\frac{\pi}{3n}\right)^{\frac{2}{3}}$). The table 1 displays the room temperature value of *n*, µ and Hall coefficient (R$_H$) for all samples. The room temperature mobility (µ) is calculated via using the relation; $\sigma = ne\mu$. A decreasing value of *n* and increasing value of µ is observed in the samples having excess Ni concentration up to 0.03. This decrease in the *n* value is counter intuitive but has been evidenced in several other studies.[11-15] Indeed, a high value of *n* will be expected because of the addition of semi-metallic FH phase in semiconducting HH matrix.[25] In literature, two theories have been proposed to explain the anomalous features achieved in these composite samples. Firstly, this decrease in *n* value could be ascribed by filtering of low energy carrier through negligible barriers created at the phase boundaries of HH and FH.[11-15] This has been schematically explained by molecular orbital theory (MOT) in Figure 4. A theoretical model by Faleev et al.[26] also proposes that an energy barrier is generated at the interface of metallic inclusion phase/semiconducting matrix phase due to band bending which ultimately plays an important role as an energy filter for the carriers to have low energy whereas the charge carriers having high energy remain unaffected. The same trends have also been discussed in the model proposed by Nolas et al.[27]. In terms of this theory; the enhancement of α in $ZrNi_{1.03}Pb_{0.38}Sn_{0.6}Bi_{0.02}$ as compared to $ZrNiPb_{0.38}Sn_{0.6}Bi_{0.02}$ HH can be linked to changes in scattering factor and reduced Fermi energy. Herein, the Seebeck coefficient (α) is given by

$$\alpha = \frac{\pi^2}{3}\frac{\kappa_B}{e}(r+\frac{2}{3})(\frac{1}{\xi}) \qquad (2)$$

Where ξ represents the reduced Fermi energy, Boltzmann constant ($k_B$) and r implies the scattering factor. The decrease in *n* value for $ZrNi_{1.03}Pb_{0.38}Sn_{0.6}Bi_{0.02}$ could therefore be hypothesized to be linked to reduction in ξ and increase r value towards overall increase in α value in contrast with HH alloys.[28] Although, our data could be qualitatively explained through Nolas's model but a quantitative analysis will be required to explain it completely. Such decreasing value of *n* leading to considerable enhancement in α has been reported in few earlier studies.[11-15]

Figure 3 (b) displays the temperature dependent σ of $ZrNi_{1+x}Pb_{0.38}Sn_{0.6}Bi_{0.02}$ (0 ≤ x ≤ 0.07) composites. A decrease in σ at 323 K was observed in $ZrNi_{1+x}Pb_{0.38}Sn_{0.6}Bi_{0.02}$ with x= 0.01 & 0.03 samples as compared to $ZrNiPb_{0.38}Sn_{0.6}Bi_{0.02}$ HH matrix. As displayed in table 1, an increasing value of carrier mobility (μ) was observed with increasing Ni concentration in $ZrNi_{1+x}Pb_{0.38}Sn_{0.6}Bi_{0.02}$ composites. This increment in μ value may be due to the formation of FH which precipitated out during phase separation process. These FH nanoclusters offer a smooth pathway for charge carriers in electronic transport which is in agreement with published reports.[11-15] Focusing on high temperature range, we find that at 773 K, the σ of the $ZrNi_{1.03}Pb_{0.38}Sn_{0.6}Bi_{0.02}$ is ~15% higher than that of the $ZrNiPb_{0.38}Sn_{0.6}Bi_{0.02}$ HH matrix (Fig. 3b). This significant increase is observed due to the mobility contribution in conductivity which seems to have dominated over and above the carrier concentration contribution leading to an enhancement in μ. In essence, increasing Ni concentration results in an enhancement in value of μ for all composite samples (Table 1). This can be interpreted as contributing to FH nano inclusions that provide facile pathway to the carriers for increasing the electronic transport. In Fig. 3(b), we also note that ZrNi$_{1.07}$Pb$_{0.38}$Sn$_{0.6}$Bi$_{0.02}$ composite sample has ~28% enhanced value of electrical conductivity as compared to ZrNiPb$_{0.38}$Sn$_{0.6}$Bi$_{0.02}$ due to combined improvement in carrier concentration and mobility. Clearly significant increase in metallicity has happened for x= 0.07 that is reflecting low value of α as well.

Insightful understanding of transport properties can also be achieved by using band bending effects as elucidated by molecular orbital theory (MOT) diagram as displayed in figure 4. We know that the band structure of FH is quite similar to HH as they have almost same composition with only difference of a partially filled band due to one extra Ni atom in FH composition. But the band gap of FH is less than HH. The deliberate excess of Ni concentration in HH is reducing the band of composite as compared to pure HH. This is presented clearly on the basis of MOT diagram to show the heterojunction formation at interface of ZrNiPb (HH) / ZrNi$_2$Pb (FH) composite. This MOT diagram relates to the present composition of ZrNi$_{1+x}$Pb$_{0.38}$Sn$_{0.6}$Bi$_{0.02}$ (0 ≤ x ≤ 0.07) composite samples in which ZrNiPb$_{0.38}$Sn$_{0.6}$Bi$_{0.02}$ represents the HH alloy and ZrNi$_2$Pb$_{0.38}$Sn$_{0.6}$Bi$_{0.02}$ corresponds to FH alloy. Figure 5 displays the relation between the temperature and power factor (PF) for all the synthesized compositions ZrNi$_{1+x}$Pb$_{0.38}$Sn$_{0.6}$Bi$_{0.02}$ (0 ≤ x ≤ 0.07) composites alloys. The sample ZrNi$_{1.03}$Pb$_{0.38}$Sn$_{0.6}$Bi$_{0.02}$ reflects maximally enhanced PF as compared to ZrNiPb$_{0.38}$Sn$_{0.6}$Bi$_{0.02}$ HH matrix. A maximum PF of 75×10$^{-4}$ W/mK$^2$ at 773 K is achieved for ZrNi$_{1.03}$Pb$_{0.38}$Sn$_{0.6}$Bi$_{0.02}$ which is ~42 % higher than that of ZrNiPb$_{0.38}$Sn$_{0.6}$Bi$_{0.02}$ HH. This high value of PF is evidently attributed to simultaneous increase in α and σ.

## 3.3 Thermal transport properties

We next discuss our results of thermal conductivity (κ) in ZrNi$_{1+x}$Pb$_{0.38}$Sn$_{0.6}$Bi$_{0.02}$ (0 ≤ x ≤ 0.07) composite samples. Figure 6 (a, b & c) elucidates the calculated temperature-dependent total thermal conductivity (κ) with corresponding lattice thermal conductivity (κ$_L$) and electronic thermal conductivity (κ$_e$) contributions for all the alloys. Herein, a significant reduction in κ was observed in ZrNi$_{1+x}$Pb$_{0.38}$Sn$_{0.6}$Bi$_{0.02}$ (0 ≤ x ≤ 0.07) composite samples as compared to ZrNiPb$_{0.38}$Sn$_{0.6}$Bi$_{0.0\,2}$ HH with simultaneous increase in PF. The variation of κ with temperature is displayed in figure 6 (a) for all the samples. In all the samples, κ value is reduced with increase in temperature which implies dominant lattice conductivity. The reduction in κ values are observed with increasing Ni content in the samples. The sample having ZrNi$_{1.07}$Pb$_{0.38}$Sn$_{0.6}$Bi$_{0.02}$ composition depicts the lowest κ value ~ 4.2 W/mK at 773 K which is ~ 25% inferior to ZrNiPb$_{0.38}$Sn$_{0.6}$Bi$_{0.02}$ HH. The decreasing value of κ could be attributed to facile scattering of phonon at various grains and phase boundaries along with their interfaces that was optimally achieved through compositional tuning approach.

The Wiedemann-Franz law was applied to obtain κ$_e$ via relation κ$_e$ = LσT, where L is the Lorentz number, electrical conductivity (σ), and temperature (T). And, κ$_L$ value was calculated by subtracting κ$_e$ value from κ value.[29] The exact value of κ$_e$ is achieved via calculating correct estimation of L with temperature. The temperature dependent L for all the samples is displayed in Fig. 7(a). Fig. 7(b) presents L vs α for the entire composite samples and also with the estimated L value through SPB-APS model[30] or can be described as a single parabolic band with acoustic phonon scattering model. This fitting also shows the degenerated limit of the Lorentz number L= 2.4 x 10$^{-8}$ W ohm K$^{-2}$.[30]

Figure 6 (b) displays the temperature variation of κ$_e$ for ZrNi$_{1+x}$Pb$_{0.38}$Sn$_{0.6}$Bi$_{0.02}$ (0 ≤ x ≤ 0.07) composite samples. An increased value of κ$_e$ is observed with increasing temperature in all composites which is consistent with σ data. We observe that κ$_L$ (Fig 6c) decreases with rising temperature, displaying similar trend as we acquire in κ. Thus, reduction in κ is mainly attributable to drastic reduction in κ$_L$. The observed reduction in κ$_L$ for ZrNi$_{1+x}$Pb$_{0.38}$Sn$_{0.6}$Bi$_{0.02}$ (0 ≤ x ≤ 0.07) composites samples are due to coherent nano-range boundaries of HH and FH which efficiently scatter mid frequency phonons in addition to longer ones at grain boundaries within these composites.[11-13]

## 3.4 Thermoelectric figure-of-merit (ZT)

Figure 8 (a) shows ZT variation with temperature in ZrNi$_{1+x}$Pb$_{0.38}$Sn$_{0.6}$Bi$_{0.02}$ (0 ≤ x ≤ 0.07) composites. Irrespective of the temperature, the ZT value for ZrNi$_{1+x}$Pb$_{0.38}$Sn$_{0.6}$Bi$_{0.02}$ composites increases with rising temperature. A substantial increase in ZT ≈ 1.3 at temperature of 773 K is achieved for ZrNi$_{1.03}$Pb$_{0.38}$Sn$_{0.6}$Bi$_{0.02}$ that is considerably higher value as compare to ZrNiPb$_{0.38}$Sn$_{0.6}$Bi$_{0.02}$ HH matrix phase (ZT ≈ 0.67 at 773 K). Thus, a synergistic increase in α ~ 12% and σ ~ 15%, and a reduction of ~ 25 % in κ, results in about 90 % enhancement in the ZT value for ZrNi$_{1.03}$Pb$_{0.38}$Sn$_{0.6}$Bi$_{0.02}$ in comparison to that of pristine HH. Whereas, ZrNi$_{1.07}$Pb$_{0.38}$Sn$_{0.6}$Bi$_{0.02}$ composite, which is semi-metallic in nature, exhibited a largest reduction in κ but could not attain large ZT value due to very low α. Hence, high concentration of FH in the composite is not profitable for enhancing ZT value. Thus, the present system of HH composite has reasonably high value of ZT as compared to other state-of-the-art n-type HH thermoelectric materials[30] including NiCoSb (ZT~ 0.4 @ 973 K), ZrNiPb (ZT~ 0.6 @ 873 K), $Nb_{0.8}Ti_{0.2}FeSb$ (ZT~ 1.0 @ 973 K), $Hf_{0.44}Zr_{0.44}Ti_{0.12}Sb_{0.8}Sn_{0.2}$ (ZT~ 0.4 @ 973 K), and $Hf_{0.8}Ti_{0.2}CoSb_{0.8}Sn_{0.2}$ (ZT~ 0.9 @ 1073 K). Moreover, these composites are free from hafnium (Hf) which makes these materials cost effective for potential thermoelectric device fabrication.[6]

### 3.5 Compatibility Factor

Thermoelectric compatibility factor (S) was estimated for all $ZrNi_{1+x}Pb_{0.38}Sn_{0.6}Bi_{0.02}$ samples. This factor is used for compatible segmentation in TE devices for waste heat recovery applications. This is defined as follows[31],

$$S = \frac{\sqrt{1+ZT}-1}{\alpha T} \qquad (3)$$

Where temperature (T) and Seebeck coefficient (α). The value of S is required for finding the best suited counterpart and to achieve maximum TE efficiency in thermoelectric device. The S value of $ZrNi_{1.03}Pb_{0.38}Sn_{0.6}Bi_{0.02}$ (~ - 2.7 $V^{-1}$ at 773K) composite is boosted up by ~17% as to the parent $ZrNiPb_{0.38}Sn_{0.6}Bi_{0.02}$ (-2.3 $V^{-1}$ at 773K) HH as shown in Fig 8 (b). The value of S for $ZrNi_{1.03}Pb_{0.38}Sn_{0.6}Bi_{0.02}$ sample calculated here is comparable to other n-type thermoelectric materials[32] which includes $Ti_{0.5}(ZrHf)_{0.5}NiSn$ (S~ -2.8 $V^{-1}$ @ 973 K), $Si_{0.78}Ge_{0.22}$ (S~ -1.7 $V^{-1}$ @ 1000 K), $Bi_2Te_3$ (S~ -2.5 $V^{-1}$ @ 573 K) MnSiSn (S~ -1.7 $V^{-1}$ @ 873 K) and $La_3Te_4$ (S~-1.3 $V^{-1}$ at 1273 K). We find that $ZrNi_{1.03}Pb_{0.38}Sn_{0.6}Bi_{0.02}$ is a suitable n-type thermoelectric compound for segmentation with other TE materials. Therefore, the two-fold benefits of high value of ZT and decent S value in $ZrNi_{1+x}Pb_{0.38}Sn_{0.6}Bi_{0.02}$ (0 ≤ x ≤ 0.07) augers well for thermoelectric device fabrication.[32]

For more clear evidence towards achieving high ZT in $ZrNi_{1+x}Pb_{0.38}Sn_{0.6}Bi_{0.02}$ composite compounds, we have estimated the theoretical conversion efficiency of thermoelectric device by using the following equation[1]:

$$\eta = \frac{\Delta T}{T_h} \frac{\sqrt{1+ZT_{avg}}-1}{\sqrt{1+ZT_{avg}}+\frac{T_c}{T_h}} \quad (4)$$

Where ΔT is the difference in temperature of hot temperature ($T_h$) and cold temperature ($T_c$ = 300 K) and $ZT_{avg}$ represents the average ZT value which as calculated by the given equation[33]:

$$ZT_{avg} = \frac{\int_{T_c}^{T_h} ZT dt}{T_h - T_c} \quad (5)$$

Here, the device theoretical conversion efficiency value for optimized composite sample $ZrNi_{1.03}Pb_{0.38}Sn_{0.6}Bi_{0.02}$ sample is estimated to be around 13.6 %. Figure 8 (c) shows a relation between temperature and average ZT value ($ZT_{avg}$) for all the compositions $ZrNi_{1+x}Pb_{0.38}Sn_{0.6}Bi_{0.02}$ (0 ≤ x ≤ 0.7) composite samples. The theoretically calculated thermoelectric device conversion efficiency (with increasing ΔT) for $ZrNi_{1+x}Pb_{0.38}Sn_{0.6}Bi_{0.02}$ composites samples is displayed in figure 8 (d). We thus establish that these compositionally tuned n-types HH based composite TE materials provide an effective pathway for TE device fabrication.[30]

## 4. Conclusion:

In conclusion, compositional tuning approach has been employed to synthesize a series of alloys having composition $ZrNi_{1+x}Pb_{0.38}Sn_{0.6}Bi_{0.02}$ (0 ≤ x ≤ 0.07). A substantially enhanced value of ZT ~ 1.3 (at 773 K) is achieved for $ZrNi_{1.03}Pb_{0.38}Sn_{0.6}Bi_{0.02}$ which is ~90% higher as compared to the pristine $ZrNi_{1.03}Pb_{0.38}Sn_{0.6}Bi_{0.02}$ HH alloy. This improved ZT is ascribed to higher power factor ($75\times10^{-4}$ W/mK$^2$ at 773 K) and reduced thermal conductivity, which is nearly ~25% lower than pristine HH alloy. It is noted that increasing Ni concentration provides multiple sizes of FH inclusion in HH matrix for facile phonon scattering which significantly reduces lattice thermal conductivity. Moreover, a low energy electron filtering at the interface results in higher value of Seebeck Coefficient leading to enhanced power factor. The thermoelectric compatibility factor (S) for x=0.03 is estimated ~ - 2.7 V$^{-1}$ at 773K which is ~17% higher than pristine composition. The composition $ZrNi_{1.03}Pb_{0.38}Sn_{0.6}Bi_{0.02}$ has revealed two-fold benefits of enhanced ZT along with appreciable S value that provide excellent pathway for efficient thermoelectric device fabrication. The theoretically calculated thermoelectric device efficiency attains a value of η ~ 13.6% for ZrNi$_{1.03}$Pb$_{0.38}$Sn$_{0.6}$Bi$_{0.02}$. Our results imply that compositional tuning approach holds great potential towards achieving new thermoelectric alloys with high conversion efficiency.

**Acknowledgement**

This work was supported by Indian Department of Science and Technology (grant no. INT/RUS/RFBR/316). Prof. S. Patnaik thanks DST-FIST for low temperature and high magnetic field facility at JNU, Delhi, India. AB acknowledges DST-SERB for NPDF award (PDF/2017/001250). VK acknowledges Russian Science Foundation (grant no. 21-12-00405). AS and AB acknowledges UGC for JRF and DSKPDF respectively.

**Conflict of interest statement**

The authors declare that they have no competing financial interests or personal relationships that could have appeared to influence the work reported in this paper.

**Data availability**

Data supporting the results and conclusions of this study are available within the article.

**Author Contributions**

**(Any of one)**

All authors contributed to the study conception and design. Material preparation, data collection and analysis were performed by AS and AB. The first draft of the manuscript was written by AB and SP and all authors commented on previous versions of the manuscript. All authors read and approved the final manuscript.

**Table Caption:**

**Table 1:** Hall measurement data for of ZrNi$_{1+x}$Pb$_{0.38}$Sn$_{0.6}$Bi$_{0.02}$ ($0 \leq x \leq 0.07$) composites samples.

**Table 1**

| Nominal composition | $R_H$ ($10^{-2}$ m$^3$ C$^{-1}$) | $n$ ($10^{20}$ cm$^{-3}$) | $\mu$ (cm$^2$V$^{-1}$s$^{-1}$) |
|---|---|---|---|
| x=0.00 | 1.50 | 4.16 | 29.4 |
| x=0.01 | 1.86 | 3.36 | 30.52 |
| x=0.03 | 1.91 | 3.26 | 41.43 |
| x=0.07 | 1.48 | 4.21 | 44.69 |

**Figure Captions:**

**Figure 1:** XRD pattern of ZrNi$_{1+x}$Pb$_{0.38}$Sn$_{0.6}$Bi$_{0.02}$ ($0 \leq x \leq 0.07$) composites samples.

**Figure 2** SEM image of best thermoelectric performing composite ZrNi$_{1.03}$Pb$_{0.38}$Sn$_{0.6}$Bi$_{0.02}$ at low-magnification. It displays *in-situ* FH precipitate indicated by white dotted circle and the matrix phase HH by solid white-square along with the EDX analysis.

**Figure 3:** (a) Seebeck coefficient and (b) electrical conductivity of ZrNi$_{1+x}$Pb$_{0.38}$Sn$_{0.6}$Bi$_{0.02}$ ($0 \leq x \leq 0.07$) composites as a function of temperature.

**Figure 4:** A representative molecular orbital theory; MOT diagram explains the development of FH (ZrNiPb) inclusion phase because of the incorporation of Ni atom in the empty tetrahedral site of HH phase (ZrNiPb).

**Figure 5:** Temperature dependent PF for all the composition of ZrNi$_{1+x}$Pb$_{0.38}$Sn$_{0.6}$Bi$_{0.02}$ ($0 \leq x \leq 0.07$) composites samples.

**Figure 6:** (a) Total thermal conductivity (b) Electronic (c) Lattice thermal conductivity of ZrNi$_{1+x}$Pb$_{0.38}$Sn$_{0.6}$Bi$_{0.02}$ ($0 \leq x \leq 0.07$) with temperature variation.

**Figure 7:** Lorentz number (L) is plotted as a function of temperature for ZrNi$_{1+x}$Pb$_{0.38}$Sn$_{0.6}$Bi$_{0.02}$ ($0 \leq x \leq 0.07$) composites samples, (b) α (T) dependent L value for ZrNi$_{1+x}$Pb$_{0.38}$Sn$_{0.6}$Bi$_{0.02}$ ($0 \leq x \leq 0.07$) composite samples.

**Figure 8:** (a) Thermoelectric figure-of-merit (ZT) variation with temperature in ZrNi$_{1+x}$Pb$_{0.38}$Sn$_{0.6}$Bi$_{0.02}$ ($0 \leq x \leq 0.07$) composites samples is plotted (b) Temperature dependent TE compatibility factor for ZrNi$_{1+x}$Pb$_{0.38}$Sn$_{0.6}$Bi$_{0.02}$ ($0 \leq x \leq 0.07$) composites samples is shown (c) Temperature dependent ZT average (ZT$_{avg}$) values for ZrNi$_{1+x}$Pb$_{0.38}$Sn$_{0.6}$Bi$_{0.02}$ ($0 \leq x \leq 0.07$) composites samples is plotted (d) Theoretically calculated thermoelectric device efficiency for ZrNi$_{1+x}$Pb$_{0.38}$Sn$_{0.6}$Bi$_{0.02}$ ($0 \leq x \leq 0.07$) is plotted as a function of ΔT.

**Figure 1:**

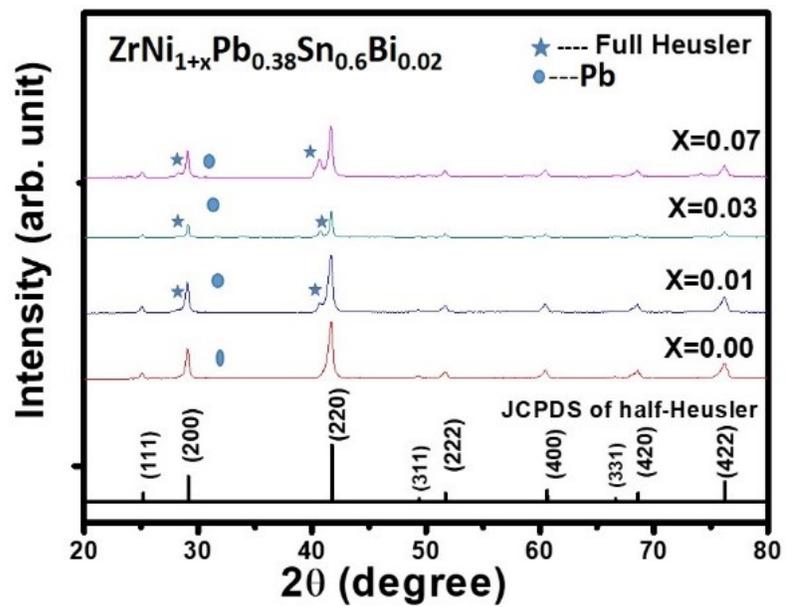

**Figure 2:**

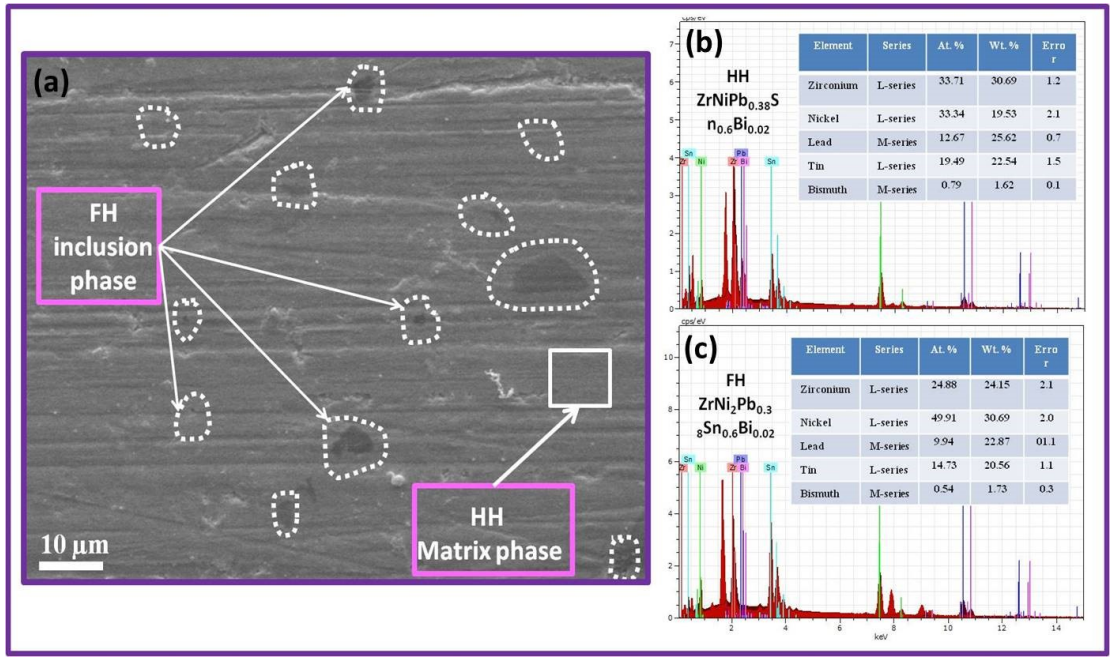

**Figure 3:**

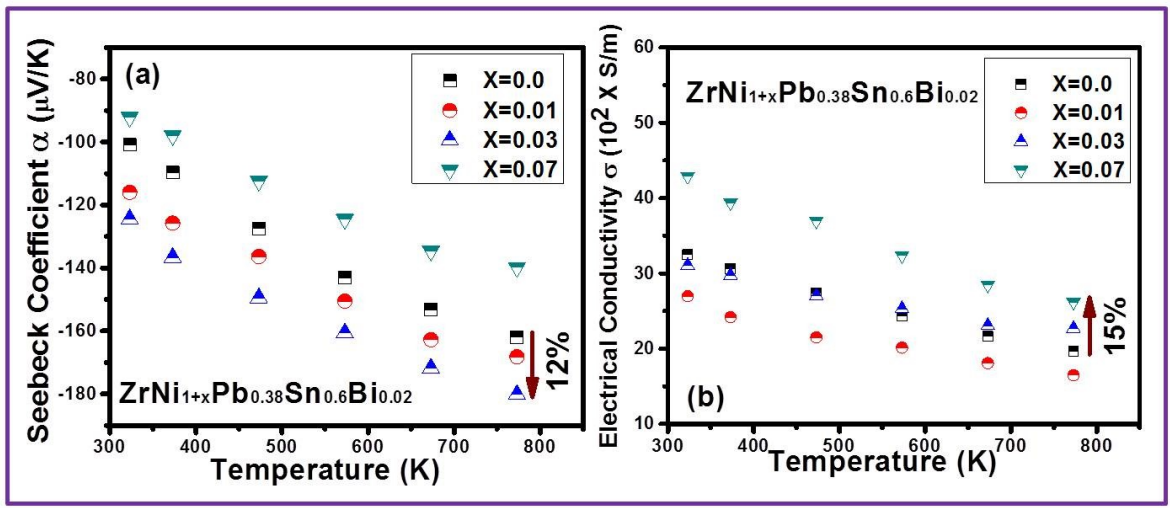

**Figure 4:**

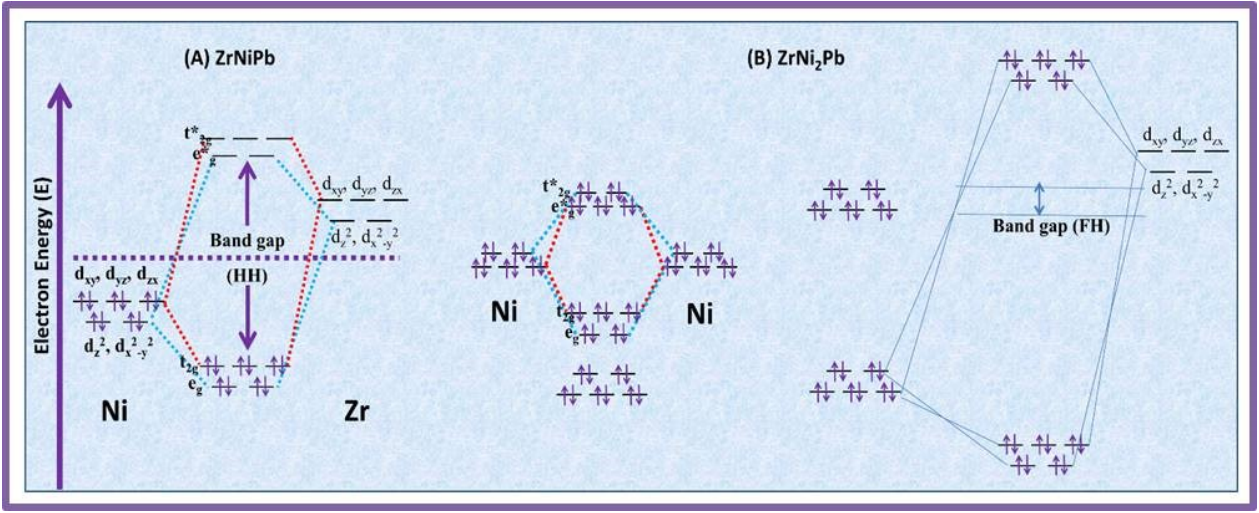

**Figure 5:**

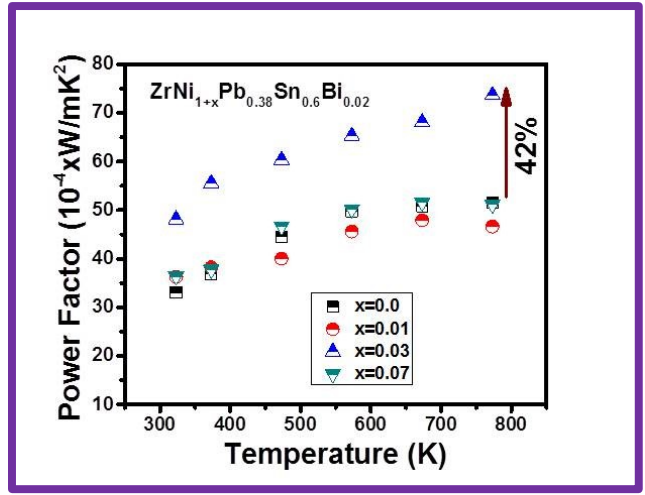

**Figure 6:**

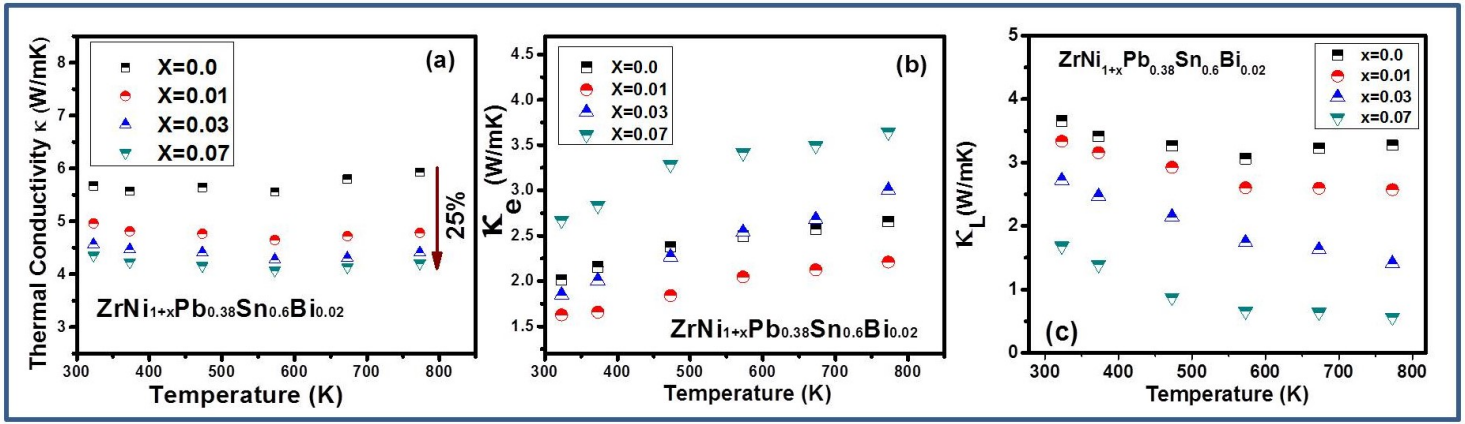

**Figure 7:**

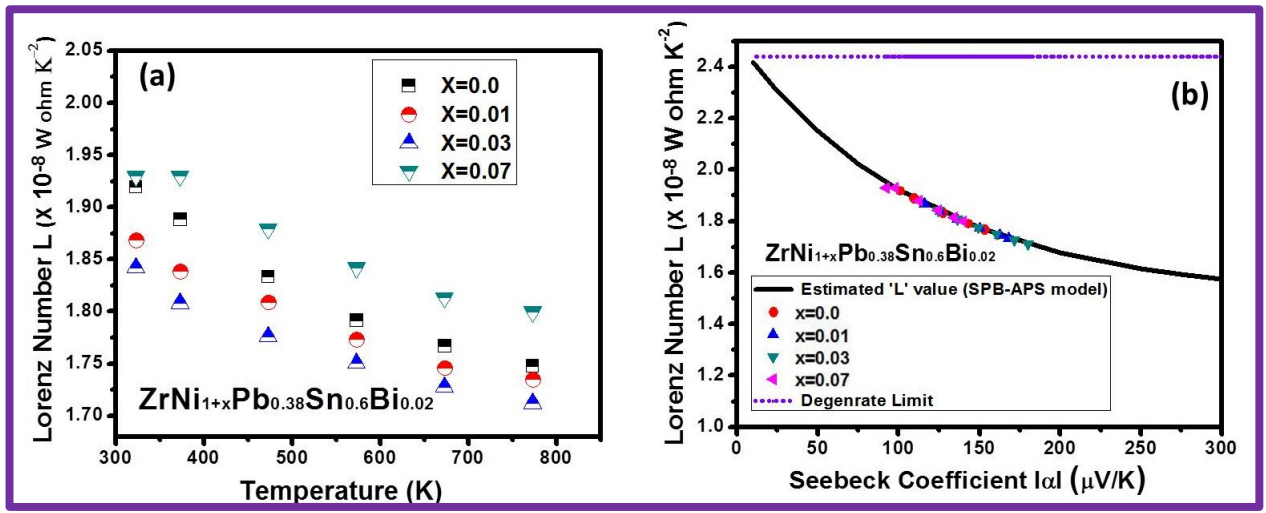

**Figure 8:**

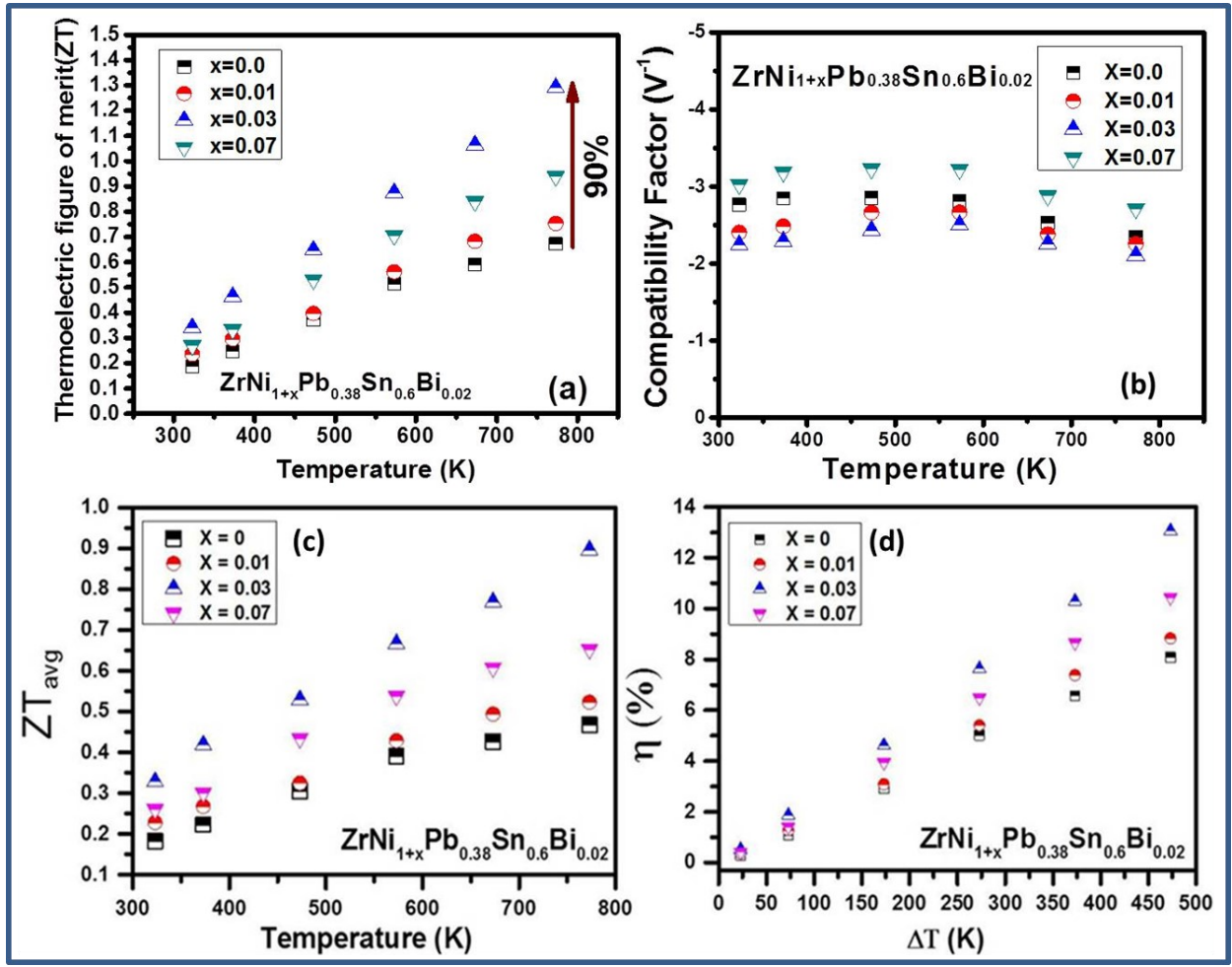